\documentclass[12pt]{JHEP3}
\usepackage{epsfig}
\usepackage{amsmath}
\usepackage{amssymb}

\newcommand{\tr}{{\rm tr}}

\newcommand{\Tr}{{\rm Tr}}
\newcommand{\Str}{{\rm Str}}
\newcommand{\n}{\notag}

\title{Graviton Propagators on Fuzzy $G/H$}

\author{Yoshihisa Kitazawa\\
High Energy Accelerator Research Organization (KEK)\\
Tsukuba, Ibaraki 305-0801, Japan\\
Department of Particle and Nuclear Physics,
The Graduate University for Advanced Studies\\
Tsukuba, Ibaraki 305-0801, Japan\\
E-mail: \email{kitazawa@post.kek.jp}}
\author{Satoshi Nagaoka\\
High Energy Accelerator Research Organization (KEK)\\
Tsukuba, Ibaraki 305-0801, Japan\\
E-mail: \email{nagaoka@post.kek.jp}}

\abstract{
We describe
closed string modes by open Wilson lines in 
noncommutative (NC) gauge theories
on compact fuzzy $G/H$ in IIB matrix model. 
In this construction the world sheet cut-off
is related to the spacetime cut-off since
the string bit of the symmetric traced Wilson line carries the 
minimum momentum on $G/H$.
We show that the two point correlation functions of 
graviton type Wilson lines in 4 dimensional 
NC gauge theories 
behave as $1/({\rm momentum})^2$.
This result suggests that 
graviton is localized on D3-brane, so we can naturally 
interpret D3-branes as our universe.
Our result is not limited to D3-brane system, and we 
generalize our analysis to  
other dimensions and even to any topology of D-brane worldvolume
within fuzzy $G/H$.}

\keywords{M(atrix) Theories, Gauge-gravity correspondence, D-branes}

\preprint{KEK-TH-1060, hep-th/0512204}

\begin{document}

\section{Introduction \label{s1}}

Noncommutative (NC) gauge theory is realized \cite{CDS,AIIKKT,Li}
by considering a NC 
background in matrix models \cite{IKKT,BFSS}.
It offers a promising possibility that it contains gravity as a quantum correction
through the UV/IR mixing effect \cite{MRS}.
In string theory,
some perturbative vacua are well-known and the relation between them are
clarified, but
there is a vast amount of moduli space to be fixed,
and we have no sufficient information to 
predict which is the nonperturbative vacuum.
Landscape is one of the major fields in the recent 
development in string theory \cite{Susskind}.
On the other hand, it is still a very fascinating idea that 
our universe is uniquely selected through the nonperturbative 
effect of string theory.
To find this mechanism, it is necessary to study 
quantum gravity from string theory point of view.

Quantum gravity itself is very difficult to study, but
in string theory, there is a duality between open string and closed
string, therefore,
we can analyze quantum gravity by using open string modes.
AdS/CFT correspondence is a well-established correspondence 
\cite{Maldacena}. But in an 
ordinary gauge theory, it might not be easy for us to probe
quantum gravity, 
since we do not keep higher tower of open string degrees of 
freedom. 
On the other hand, NC gauge theories may include
such open string modes since they are essentially matrix models.
In this sense,
new effects of quantum gravity might be seen 
in the quantum corrections of NC gauge theory.
Then, what kind of phenomena is included in this effects?
One of the possibility which we discuss in this paper is
that the 4 dimensional quantum gravity is realized in 4 dimensional
NC gauge theory.
Our scenario is similar to the brane world scenario \cite{RS}, 
which explains the localization of gravity on D-brane.
We suggest that NC gauge theories provide a localization of gravity
on D-branes. 
Our goal is to derive $1/({\rm momentum})^2$ dependence 
of massless graviton propagators
in NC gauge theories.

In section \ref{s21},
we briefly review the open Wilson lines in NC gauge theories on $S^2\times S^2$. 
By considering the regularized space,
we can consider the large but finite $N$ system which 
serves us as a gauge invariant regularization. 
In section \ref{s22},
two point function of open Wilson lines
 which couple to the massless graviton mode
is calculated. The tensor structure of Wilson line correlators (WLC), 
which depends on the 
isometry of $S^2 \times S^2$, is constructed in section \ref{s23}.
In section \ref{s31}, we show that
the essential part of the correlator, 
which we explain there in detail, does not depend on our
choice of $G/H$. 
In section \ref{s32}, we calculate two point function of WLC on
 another homogeneous space, $CP^2$,
which has higher symmetry than $S^2\times S^2$.
In section \ref{s33}, we generalize our result to other dimensions, 
for example, $S^2\times S^2\times S^2$.
We conclude in section \ref{s4} with discussions.

\section{Wilson line correlators in noncommutative gauge theory \label{s2}}

Noncommutative (NC) gauge theories on compact homogeneous spaces 
can be 
constructed from IIB matrix model. They have been investigated in
\cite{KTT2,fuzS2,fuzS2S2,KTT1,fuzS2S2S2,fuzCP2}.
By considering the compact homogeneous space,
we can deal with a large but 
finite $N$ system, which enables us to investigate 
non-perturbative questions.
It thus serves us as a nonperturbative and gauge invariant regularization of 
NC gauge theory.
The bosonic part of the action of IIB matrix model is written as
\begin{align} 
S= - \frac{1}{4} \tr [A_\mu,A_\nu]^2 \ ,
\label{IIBac}
\end{align}
where $A_\mu$ are $N \times N$ Hermitian matrices and $\mu$ and $\nu$
run over 0, $\cdots$, 9. 
The equation of motion is obtained as
\begin{align}
[A_\mu,[A_\mu,A_\nu]]=0 \ .
\end{align}

NC gauge theory is obtained by expanding matrices
around the NC backgrounds.
We will denote the NC gauge field $a_\mu$ around the background $p_\mu$ as
\begin{align} 
\label{f1}
A_\mu =f_{\alpha} (p_\mu+a_\mu) \ ,
\end{align}
where $f_\alpha$ is a scale factor. 
When we consider the action (\ref{IIBac}) with a Myers term \cite{Myers} as 
\begin{align} \label{f2}
{i\over 3}f_{\mu\nu\rho} A_{\mu} [A_\nu,A_\rho] \ ,
\end{align}
we can identify a scale factor $f$ in (\ref{f1}) with
a coefficient $f$ in (\ref{f2}).
In this sense, 
the index $\alpha$ labels
the representation of a fuzzy homogeneous space \cite{Mathom}.
Alternatively such a space may be realized as a quantum solution
\cite{fuzS2S2}.
Although supersymmetry is softly broken in either case,
the leading behavior of the correlators is constrained by
SUSY.

\subsection{Feynman rule of noncommutative gauge theory on
  $S^2 \times S^2$ \label{s21}}

Let us briefly describe the 
Feynman rule of NC gauge theory on $S^2 $ with $U(1)$ gauge group. 
We will generalize the rule to $S^2 \times S^2$ background 
with $U(n)$ gauge group later.
We follow the notation in \cite{KTT2}.

We expand matrices in terms of matrix spherical harmonics as
\begin{align}
A^\mu = f_{S^2} (p^\mu + \sum_{jm} a_{jm}^{\mu}Y_{jm}) \ ,
\end{align}
where the representation $Y_{jm}$ is adopted as
\begin{align}
&(Y_{jm})_{ss'}=(-1)^{l-s}
\left(\begin{array}{ccc}
l&j&l_{} \\
-s&m&s'
\end{array}\right)
\sqrt{2j+1} .
\end{align}
$p_\mu$ can be identified with the angular momentum operator
in the spin $l$ representation.
The normalization is defined as
\begin{align}
\mbox{Tr}\;Y_{j_1m_1}Y_{j_2m_2}=(-1)^{m_1}\delta_{j_1,j_2}
\delta_{m_1,-m_{2}} \ .
\end{align}
The cubic vertex of matrix spherical harmonics is written as
\begin{align} \label{3pv}
\begin{picture}(0,0)
\put(5,0){\circle{20}}
\put(15,0){\line(1,0){10}}
\put(-15,5){\line(1,0){10}}
\put(-15,-5){\line(1,0){10}}
\put(-14,10){$Y_{j_2}$}
\put(-14,-16){$Y_{j_1}$}
\put(15,4){$Y_{j_3}$}
\end{picture} \hspace*{1cm}\n
=\Tr[Y_{j_1m_1}Y_{j_2m_2}Y_{j_3m_3}]
&=(-1)^{2l}\sqrt{(2j_1+1)(2j_2+1)(2j_3+1)}\\
&\times\left(\begin{array}{ccc}
j_1&j_2&j_3\\
m_1&m_2&m_3
\end{array}\right)
\left\{\begin{array}{ccc}
j_1&j_2&j_3\\
l&l&l
\end{array}\right\} \ ,
\end{align}
where we adopt the notation of $(3j)$ and $\{6j\}$ symbols in \cite{Edm}.
The propagators of the NC gauge field $a_{jm}^\mu$ are read 
from the action as
\begin{align}
\langle \; a^{\mu}_{j_1m_1}a^{\nu}_{j_2m_2}\;\rangle =
\frac{1}{f_{S^2}^{4}}\;\frac{(-1)^{m_1}}{j_1(j_1+1)}\;
\delta^{\mu\nu}\delta_{j_1j_2}\delta_{m_1-m_{2}}  \ .
\end{align}

Next, let us introduce Wilson lines in NC gauge theory \cite{IIKK} 
\footnote{The large momentum limit of Wilson line correlators
is discussed in \cite{Gross,DhKhe}.} on $S^2$.
They are constructed by the trace of polynomial of matrices as
\begin{align}
y_{jm}^{\alpha_1,\alpha_2,\cdots,\alpha_j}TrA_{\alpha_1}A_{\alpha_2}\cdots A_{\alpha_j}
A_{i_1}A_{i_2}\cdots A_{i_k} \ .
\end{align}
$\alpha=7,8,9$ denote the dimensions where $S^2$ is embedded.
$y_{jm}^{\alpha_1,\alpha_2,\cdots,\alpha_j}$ denotes
a totally symmetric traceless tensor which corresponds to the spin $j $
representation of $SU(2)$.
The background $p_\mu$ consists of angular momentum operators in spin 
$l$ representation.
In our expansion of $A_\mu$ around the background $p_\mu$,
the leading term of the Wilson line is written as
\begin{align}
f_{S^2}^{j+k}
y_{jm}^{\alpha_1,\alpha_2,\cdots,\alpha_j}
\Tr p_{\alpha_1}p_{\alpha_2}\cdots p_{\alpha_j}
{\cal O}_1 \cdots {\cal O}_k \ ,
\end{align}
where ${\cal O}$ is a field around the background $p_\mu$.
We define ${\cal Y}_{jm}$ as
\begin{align}
{\cal Y}_{jm}&\equiv y_{jm}^{\alpha_1,\alpha_2,\cdots,\alpha_j}
p_{\alpha_1}p_{\alpha_2}\cdots p_{\alpha_j} \ .
\end{align}
We will focus on the highest weight states of $SU(2)$, 
therefore, we also define ${\cal Y}_{j}$ as
\begin{align}
y_{j,j}\Tr(p_{+})^{j} {\cal O}_1 \cdots {\cal O}_k 
=\Tr{\cal Y}_{j}{\cal O}_1 \cdots {\cal O}_k \ ,
\end{align}
where $p_+\equiv p_7+ip_8$. 

Using these Feynman rules, we find that 
there are planar and non-planar contribution in the 
two point function of $\Tr{\cal Y}_{j}{\cal O}_1 {\cal O}_2$
at the leading order,
\begin{align}
& {1\over {2l+1}}\langle \Tr{\cal Y}_{j}{\cal O}_1 {\cal O}_2
\Tr {\cal O}_2^\dagger {\cal O}_1^\dagger {\cal Y}_{j}^{\dagger}\rangle
\ \n \\
& =
\hspace*{1.5cm}
\begin{picture}(0,0)
\put(-40,0){\line(1,0){5}}
\put(-25,0){\circle{20}}
\put(5,0){\circle{20}}
\put(15,0){\line(1,0){5}}
\put(-15,5){\line(1,0){10}}
\put(-15,-5){\line(1,0){10}}
\put(-14,8){${\cal O}_1$}
\put(-14,-15){${\cal O}_2$}
\put(-45,4){${\cal Y}$}
\end{picture}\hspace*{8mm}
=\langle j|{1\over P_1^2P_2^2}|j\rangle_{\rm p} \ , \n \\
& {1\over {2l+1}}\langle\Tr{\cal Y}_{j}{\cal O}_1 {\cal O}_2
\Tr {\cal O}_1^\dagger {\cal O}_2^\dagger {\cal Y}_{j}^{\dagger}\rangle \n \\
& =\hspace*{1.5cm}
\begin{picture}(0,0)
\put(-40,0){\line(1,0){5}}
\put(-25,0){\circle{20}}
\put(5,0){\circle{20}}
\put(-10,0){\line(1,0){5}}
\put(-15,5){\line(1,0){10}}
\put(-15,-5){\line(1,0){10}}
\put(-14,8){${\cal O}_1$}
\put(-14,-15){${\cal O}_2$}
\put(-45,4){${\cal Y}$}
\end{picture}\hspace*{8mm}
=\langle j|{1\over P_1^2P_2^2}|j\rangle_{\rm np} \ ,
\end{align}
where
\begin{align}
P_i^{\mu}{\cal Y}_{j_{i'}m_{i'}}&\equiv  [p^{\mu},{\cal Y}_{j_{i'}m_{i'}}]
\delta_{ii'} \ .
\end{align}
The planar and nonplanar part of the correlation function on $S^2$ is 
given by
\begin{align}
\langle j|X|j\rangle_p
&={1\over f_{S^2}^8 (2l+1)}\sum_{j_2,j_3,m_2,m_3}
\Psi_{123}^*X \Psi_{123} \ ,\n \\
\langle j|X|j\rangle_{np}
&={1\over f_{S^2}^8 (2l+1)}\sum_{j_2,j_3,m_2,m_3}
\Psi_{132}^*X \Psi_{123} \ ,\n \\
{\rm where} \hspace*{1cm}\Psi_{123}&\equiv 
\Tr {\cal Y}_{j_3m_3}{\cal Y}_{j_2m_2}{\cal Y}_{j}.
\end{align}

Now, let us formulate the Wilson line correlators on $S^2 \times S^2$
with $U(n)$ gauge group.
The construction is the simple extension 
of the correlators on $S^2$.
We expand matrices in terms of the
tensor product of matrix spherical harmonics as
\begin{align}
A_{\mu}=f_{\rm S^2\times S^2}
(p_{\mu}+\sum_{jmpq}a^{\mu}_{jmpq}Y_{jm}\otimes Y_{pq}) \ ,
\end{align}
where 
\begin{align}
&p_{\mu}=j_{\mu}\otimes 1  ~~(\mu=4,5,6) \ ,\n \\
&p_{\mu}=1\otimes \tilde{j}_{\mu} ~~(\mu=7,8,9) \ .
\end{align}
In this section, we consider only the $S^2 \times S^2$ manifold,
therefore, from now on, 
we denote $f_{S^2 \times S^2}$ as $f$.
The summations over $j$ and $p$ run up to
$j=2l$ and $p=2l$ respectively.
We consider NC gauge theory with $U(n)$ gauge group, so $N=n(2l+1)^2$.
The propagators are written as
\begin{align}
\langle\; a^{\mu}_{j_1m_1p_1q_1}a^{\nu}_{j_2m_2p_2q_2}\;\rangle&=&
\frac{1}{f^{4}}\;
\frac{(-1)^{m_1+q_1}}{j_1(j_1+1)+p_1(p_1+1)}\;
\delta^{\mu\nu}\delta_{j_1j_2}\delta_{p_1p_2}
\delta_{m_1-m_{2}}\delta_{q_1-q_{2}} \ .
\end{align}
We define the normalization as
\begin{align}
\Tr Y_{j_1m_1p_1q_1}Y_{j_2m_2p_2q_2}=n (-1)^{m_1}\delta_{j_1j_2}
\delta_{m_1-m_{2}} \delta_{p_1p_2} \delta_{q_1-q_2} \ .
\end{align}
The planar and nonplanar part of the correlation function on 
$S^2 \times S^2$ are given by
\begin{align}
\langle j,p |X|j,p \rangle_p
&={n^3\over f^8 N}\sum_{j_2,j_3,m_2,m_3} \sum_{p_2,p_3,q_2,q_3}
\Psi_{123}^*X \Psi_{123} \ ,\n \\
\langle j,p|X|j,p\rangle_{np}
&={n^3\over f^8 N}\sum_{j_2,j_3,m_2,m_3} \sum_{p_2,p_3,q_2,q_3}
\Psi_{132}^*X \Psi_{123} \ ,\n \\
{\rm where} \hspace*{1cm}\Psi_{123}&\equiv 
\Tr {\cal Y}_{j_3m_3p_3q_3}{\cal Y}_{j_2m_2p_2q_2}{\cal Y}_{j,p} \ .
\end{align}
The leading terms of the Wilson lines in the
highest weight state representation 
of $SU(2) \times SU(2)$ are written as
\begin{align}
y_{j,j}y_{p,p}\Tr(p_{+})^{j}({\tilde{p}_{+}})^p 
{\cal O}_1 \cdots {\cal O}_k 
=\Tr{\cal Y}_{j,p}{\cal O}_1 \cdots {\cal O}_k \ .
\end{align}
Finally, we define $\lambda \equiv \frac{n^2}{f^4 N}$, which
is identified with 't Hooft coupling.

\subsection{Two point correlation function of massless graviton mode
\label{s22}}

The relation between straight Wilson line operators and 
fields in the massless
supergravity multiplet is clarified in \cite{vertex}\cite{ITU}.
In this section, we investigate the two point correlators 
of a massless graviton mode. 
The vertex operators which couple to the graviton in type IIB matrix
model are written as
\begin{align}
\Str exp(ik\cdot A)
([A^{\rho},A^{\mu}][A^{\rho},A_{\nu}]
+{1\over 2}\bar{\psi}\Gamma^{(\nu}[A^{\mu )},\psi])
h_{\nu\mu}\n \\
+ {1\over 2}\Str exp(ik\cdot A) \bar{\psi}\Gamma^{\rho\beta(\nu}\psi
[A^{\mu )},A_{\beta}]\partial_{\rho}h_{\nu\mu} \ ,
\end{align}
where the symbol $\Str$ implies that the ordering of the matrices is
defined through the symmetric trace.
$(\mu,\nu)$ implies that the Lorentz indices are symmetrized.
In analogy with this operator, we may introduce the 
Wilson line operator in NC gauge theory on 
$S^2 \times S^2$ as
\begin{align}
\Str {\cal Y}_{j, p }(A) ([A_\rho,A_\mu][A_\rho,A_\nu]
+{1\over 2}\bar{\psi}\Gamma^{(\nu}[A^{\mu )},\psi]) \ .
\end{align}
The symmetric trace of the operators 
on compact space may be defined as 
\begin{align}
\Str  (p_+)^{j} (\tilde{p}_+)^{p}{\cal O}_1 {\cal O}_2
&\equiv \frac{1}{j}
 \Tr \sum_{j_1=0}^j 
(p_+)^{j_1} (\tilde{p}_+)^{p_1}{\cal O}_1 
(p_+)^{j-j_1} (\tilde{p}_+)^{p-p_1}
{\cal O}_2 \ , \n \\
&{\rm where} \hspace*{5mm}
p_1 \sim \frac{p}{j}j_1 \ ,
\end{align}
which is a natural extension of the symmetric trace 
in the flat noncommutative space. 
$p_1$ is an integer nearest to $j_1p/j$.
Although supersymmetry is softly broken at the scale
where the manifold is curved, it will not affect the leading behavior
of the correlators with respect to the large $N$ limit.

The leading term of the Wilson line is written as
\begin{align}
\Str {\cal Y}_{j, p} ([p_\rho,a_\mu]-[p_\mu,a_\rho])
([p_\rho,a_\nu]-[p_\nu,a_\rho]) \equiv
\Str {\cal Y}_{j, p} f_{\rho\mu} f_{\rho\nu}
\ .
\end{align}
where we define $f_{\rho\mu}\equiv [p_\rho,a_\mu]-[p_\mu,a_\rho]$.
Note that there are other terms in the expansion, for example,
\begin{align}
\Str {\cal Y}_{j, p} [a_\rho,a_\mu] [a_\rho,a_\nu] \ .
\end{align}
But these terms are of higher orders with respect to the 't Hooft coupling 
$\lambda$.
The two point function of the Wilson line operator which couples to 
graviton is written as
\begin{align} \label{tensor}
\langle\; \Str {\cal Y}_{j, p} f_{\rho\mu} f_{\rho\nu}
\Str f_{\rho'\mu'}^\dagger f_{\rho'\nu'}^\dagger {\cal Y}_{j, p}^\dagger
\;\rangle \ .
\end{align}
First, we simplify the correlators in such a way that
\begin{align}
f_{\rho\mu}\rightarrow f_1=[p_{\rho},a_{\mu}],
~~f_{\rho\nu}\rightarrow f_2=[p_{\rho},a_{\nu}] \ .
\end{align}
This substitution is useful to understand the essential feature of the
correlators.
We will present the complete calculation of the correlators
in section \ref{s23}. 

In this way, we obtain
\begin{align}\label{strsum}
&\langle\; \Str {\cal Y}_{j, p} f_{1} f_{2}
\Str f_{2}^\dagger f_{1}^\dagger {\cal Y}_{j, p}^\dagger
\;\rangle \n \\
&=y_j^2 y_p^2 \langle\; \Str (p_+)^{j} (\tilde{p}_+)^{p} f_1 f_2
\Str f_2^\dagger f_1^\dagger (p_-)^{j} (\tilde{p}_-)^{p}
\;\rangle \n \\
&=  \frac{y_j^2y_p^2}{j^2} \sum_{j_1=0}^j  \sum_{j_2=0}^j 
 \langle\; \tr (p_+)^{j_1} (\tilde{p}_+)^{p_1}
 f_1 (p_+)^{j-j_1} (\tilde{p}_+)^{p-p_1} f_2 \n \\
&\hspace*{4cm}\tr f_2^\dagger (p_-)^{j-j_2} (\tilde{p}_-)^{p-p_2} f_1^\dagger 
(p_-)^{j_2} (\tilde{p}_-)^{p_2}
\;\rangle \n\\
&=\frac{y_j^2y_p^2}{j^2 }
 \sum_{j_1=0}^j  \sum_{j_2=0}^j 
(y_{j_1} y_{p_1} y_{j_2} y_{p_2}
y_{j-j_1} y_{p-p_1} y_{j-j_2} y_{p-p_2})^{-1}
\hspace*{1.7cm}
\begin{picture}(0,0)
\put(-40,0){\line(1,0){5}}
\put(-5,0){\line(1,0){5}}
\put(20,0){\line(1,0){5}}
\put(55,0){\line(1,0){5}}
\put(-20,0){\circle{30}}
\put(40,0){\circle{30}}
\put(-5,5){\line(1,0){30}}
\put(-5,-5){\line(1,0){30}}
\put(-15,0){${\cal Y}$}
\put(5,10){$f_1$}
\put(-45,4){${\cal Y}$}
\put(5,-15){$f_2$}
\end{picture}\hspace*{8mm} 
\hspace*{15mm} \ ,
\end{align}
where we denote
\begin{align}
\begin{picture}(0,0)
\put(-40,0){\line(1,0){5}}
\put(-5,0){\line(1,0){5}}
\put(20,0){\line(1,0){5}}
\put(55,0){\line(1,0){5}}
\put(-20,0){\circle{30}}
\put(40,0){\circle{30}}
\put(-5,5){\line(1,0){30}}
\put(-5,-5){\line(1,0){30}}
\put(-15,0){${\cal Y}$}
\put(5,10){$f_1$}
\put(-45,4){${\cal Y}$}
\put(5,-15){$f_2$}
\end{picture} \hspace*{2.2cm}
\equiv
\langle\; \tr {\cal Y}_{j_1, p_1} f_{1} {\cal Y}_{j-j_1,p-p_1} f_{2}
\tr f_{2}^\dagger {\cal Y}^\dagger_{j-j_2,p-p_2}
f_{1}^\dagger {\cal Y}_{j_2, p_2}
\;\rangle \ .
\end{align}
\vspace*{-2mm}

Before proceeding further, let us show
a property of the operator $f_i$, which helps us to 
perform the calculation:
\begin{align} \label{comp0}
<f_{1}f_{1}^\dagger> &\sim <[p_\rho ,a_\mu][ a_\nu^\dagger, 
p_\rho]> \n \\
&\sim \sum_{jmpq} {\cal Y} P^2 \frac{1}{P^2} ({\cal Y}^\dagger)
\delta_{\mu\nu}\sim 
\sum_{jmpq} {\cal Y}({\cal Y}^\dagger)\delta_{\mu\nu} \ .
\end{align}
Thus,
we can use the completeness condition:
\begin{align}\label{comp1}
\sum_{jmpq} ({\cal Y})_{ab} ({\cal Y}^\dagger)_{cd} 
=\delta_{ad} \delta_{bc} \ ,
\end{align}
when we sum over the internal momenta.
Here $a,b,c$ and $d$ are indices of matrices.
Note that this property 
does not depend on the choice of the basis.

Now, let us resume the calculation of (\ref{strsum}).
We substitute the results (\ref{comp0}) and (\ref{comp1}) into (\ref{strsum}) as
\begin{align}
&\langle\; \Str {\cal Y}_{j, p} f_1 f_2
\Str f_2^\dagger f_1^\dagger {\cal Y}_{j, p}^\dagger
\;\rangle  \n \\
&=\frac{n^2  }{ j^2} \sum_{j_1=0}^j  \sum_{j_2=0}^j
{y_{j}^{2}y_{p}^{2}\over
y_{j_1} y_{p_1}y_{j_2} y_{p_2}y_{j-j_1} y_{j-p_1}y_{j-j_2} y_{p-p_2}}
\n \\
&\times \tr  {\cal Y}_{j_1p_1}{\cal Y}_{j_2p_2}^{\dagger}
\tr  {\cal Y}_{j-j_1,p-p_1}{\cal Y}_{j-j_2,p-p_2}^{\dagger}
 \n \\
&=  \frac{n^2 }{ j^2} \sum_{j_1=0}^j  \sum_{j_2=0}^j
{y_{j}^{2}y_{p}^{2}\over
y_{j_1} y_{p_1}y_{j_2} y_{p_2}y_{j-j_1} y_{j-p_1}y_{j-j_2} y_{p-p_2}}
\delta_{j_1-j_2,0} \delta_{p_1-p_2,0} \n \\
&=  \frac{n^2 }{ j^2} \sum_{j_1=0}^j 
({y_jy_p\over  y_{j_1} y_{p_1}y_{j-j_1} y_{p-p_1}})^{2} \n \\
&\equiv  \frac{n^2 }{ j^2} \sum_{j_1=0}^j 
B_{j_1,j-j_1}^2 B_{p_1,p-p_1}^2 \ .
\end{align}
where we have introduced the separating function 
$B_{j_1,j-j_1}=\frac{y_j}{y_{j_1}y_{j-j_1}}$.

$B_{j_1,j-j_1}$ depends on 
a homogeneous space $G/H$ we consider.
As $(p_+)^j=(p_+)^{j_1}(p_+)^{j-j_1}$ leads to 
${\cal Y}_j=B_{j_1,j-j_1}{\cal Y}_{j_1} {\cal Y}_{j-j_1}$,
\begin{align}
B^{-1}_{j_1,j-j_1}
&=\tr {\cal Y}_j^\dagger {\cal Y}_{j_1} 
{\cal Y}_{j-j_1} \n \\
&=(-1)^{2l} \sqrt{(2j+1)(2j_1+1)(2(j-j_1)+1)} \n \\
& \times\left(\begin{array}{ccc}
j&j_1&j-j_1\\
j&-j_1&-j+j_1
\end{array}\right)
\left\{\begin{array}{ccc}
j&j_1&j-j_1\\
l&l&l
\end{array}\right\}  \ ,
\end{align}
in the case of $S^2 \times S^2$.
$(3j)$ symbol is calculated as
\begin{align}
\left(\begin{array}{ccc}
j&j_1&j-j_1\\
j&-j_1&-j+j_1
\end{array}\right)=\sqrt{\frac{1}{2j+1}} \ ,
\end{align}
while $\{6j\}$ symbol is
\begin{align}
\left\{\begin{array}{ccc}
j&j_1&j-j_1\\
l&l&l
\end{array}\right\}
\sim \sqrt{\frac{1}{2l}} 
\left(\begin{array}{ccc}
j&j_1&j-j_1\\
0&0&0
\end{array}\right) \ ,
\end{align}
when $l>>1$ \cite{VMK}. Using the Stirling formula $n! \sim \sqrt{2 \pi n}
n^n e^{-n}$, we obtain
\begin{align}
\left(\begin{array}{ccc}
j&j_1&j-j_1\\
0&0&0
\end{array}\right) 
&=(-1)^j \sqrt{\frac{(2j-2j_1)!(2j_1)!}{(2j+1)!}} \frac{j!}{(j-j_1)!j_1!}
\n \\
&\sim \left(\frac{1}{4 \pi j (j-j_1)j_1}\right)^{1/4} \ .
\end{align}
In this way, $B_{j_1,j-j_1}$ is obtained as
\begin{align}
B_{j_1,j-j_1}^2\sim l \sqrt{\frac{ \pi j}{j_1(j-j_1)}} 
\ .
\end{align}
for $j, j_1, j-j_1 >>1$.

When the momenta are equally shared: $j=p=K/2$,
Wilson line correlator (\ref{strsum}) is found as
\begin{align}
&\langle\; \Str {\cal Y}_{j,p} f_1 f_2
\Str f_2^\dagger f_1^\dagger {\cal Y}_{j,p}^\dagger
\;\rangle  \n \\
&\sim N\frac{n \pi }{K^2} \log{K^2}   \ .
\label{logk}
\end{align}
Thus, we have obtained $1/K^2$ dependence
except for the $\log K$ factor.
When we consider the correlators with $j\neq p$,
they do not exhibit $SO(4)$ symmetry.
This undesirable feature may be overcome if we consider 
the space with higher symmetry.
In fact, we will find that there are no $\log$ factor nor
directional asymmetry in the 
$CP^2$ space in section \ref{s32}.

\subsection{Ward identity for Wilson line correlators 
and tensor structure \label{s23}}

In the preceding sub-section, we have found that the graviton two point
function behaves as that of a propagator of massless field ($1/K^2$).
In this sub-section, we present the complete calculation including the 
fermionic contribution.
We will show that the tensor structure of the Wilson line 
correlators is consistent with Ward identity.

The two point function of (\ref{tensor}) is written as
\begin{align}
&\langle\; \Str {\cal Y}_{j, p} f_{\rho\mu} f_{\rho\nu}
\Str f_{\rho'\nu'}^\dagger f_{\rho'\mu'}^\dagger {\cal Y}_{j, p}^\dagger
\;\rangle \n \\
& =\langle\; \Str {\cal Y}_{j, p} ([p_\rho,a_\mu]-[p_\mu,a_\rho])
([p_\rho,a_\nu]-[p_\nu,a_\rho]) \n \\
& \Str ([p_{\rho'}^\dagger,a_{\nu'}^\dagger] 
-[p_{\nu'}^\dagger,a_{\rho'}^\dagger] )
([p_{\rho'}^\dagger,a_{\mu'}^\dagger] -
[p_{\mu'}^\dagger,a_{\rho'}^\dagger])
{\cal Y}_{j, p}^\dagger
\;\rangle \ ,
\end{align}
where we focus on the leading terms of the 't Hooft coupling $\lambda$.
Two propagators in this correlator carry almost
the same angular momenta since 
the external angular momentum is assumed to be very small 
compared to the internal angular momenta of the cut-off scale.
It is because the correlator is quartically divergent in power counting.
Therefore, we do not distinguish the two propagators 
and as a result, we obtain the following expression:
\begin{align} 
 \langle\; &\Str {\cal Y}_{j, p} ([p_\rho,a_\mu]-[p_\mu,a_\rho])
([p_\rho,a_\nu]-[p_\nu,a_\rho]) \n \\
& \Str ([p_{\rho'}^\dagger,a_{\nu'}^\dagger] 
-[p_{\nu'}^\dagger,a_{\rho'}^\dagger] )
([p_{\rho'}^\dagger,a_{\mu'}^\dagger] -
[p_{\mu'}^\dagger,a_{\rho'}^\dagger])
{\cal Y}_{j, p}^\dagger
\;\rangle
\n \\
=  & \frac{1}{K^2} \sum_{K_1, K_2} \sum_{a,b}
B_{K_1,K-K_1}^2 B_{K_2,K-K_2}^2 \n \\
\tr &{\cal Y}_1{\cal Y}_a{\cal Y}_{1'}{\cal Y}_b 
(\frac{1}{P^2})^2 \Big(
2(d-2) P^\mu P^{\mu'} P^{\nu} P^{\nu'}  \n \\
+&P^2 (2 P^{\mu}P^{\nu} \delta_{\mu'\nu'}+
2 P^{\mu'}P^{\nu'} \delta_{\mu\nu} 
 - P^{\mu}P^{\mu'} \delta_{\nu\nu'}
- P^{\nu}P^{\nu'} \delta_{\mu\mu'}
-
P^{\mu}P^{\nu'} \delta_{\mu'\nu}-
P^{\mu'}P^{\nu} \delta_{\mu\nu'}
)\n \\
 +&P^4 (\delta_{\mu\mu'}\delta_{\nu\nu'}+
\delta_{\mu\nu'}\delta_{\mu'\nu}) \Big) 
\tr {\cal Y}_b^\dagger {\cal Y}_{2'}^\dagger {\cal Y}_a^\dagger {\cal Y}_2^\dagger \ , 
\label{tens1} 
\end{align}
where 
$d(10)$ is a number of bosonic matrices.
$K_1$ and $K_2$ specify the phase structure of the left and right
sides of the symmetric trace. 
${\cal Y}_{i(i')} (i=1,2)$ are related to ${\cal Y}_{j,p}$ 
as ${\cal Y}_{j,p}=B_{K_i,K-K_i} {\cal Y}_i {\cal Y}_{i'}$.

On $ S^2 \times S^2$,
$\sum_{a,b} \tr {\cal Y}_1{\cal Y}_a {\cal Y}_{1'} {\cal Y}_b
\tr {\cal Y}_b^\dagger {\cal Y}_{2'}^\dagger {\cal Y}_a^\dagger {\cal Y}_2^\dagger
=\delta_{j_1-j_2,0} \delta_{p_1-p_2,0}$.
We have evaluated the essential part of the correlators in the 
preceding sub-section as
\begin{align}
\frac{1}{K^2}\sum_{K_1, K_2} \sum_{a,b} B_{K_1,K-K_1}^2 B_{K_2,K-K_2}^2
\tr {\cal Y}_1{\cal Y}_a {\cal Y}_{1'} {\cal Y}_b
\tr {\cal Y}_b^\dagger {\cal Y}_{2'}^\dagger {\cal Y}_a^\dagger {\cal Y}_2^\dagger 
=\frac{1}{K^2} \sum_{K_1} B_{K,K-K1}^4 \ .
\end{align}
We will focus on the tensor structure of the correlators in this sub-section.

The leading contribution of 
the fermionic part of the Wilson line correlators is obtained as
\begin{align}
&\langle\; \Str {\cal Y}_{j,p} \frac{1}{2} \bar{\psi} \Gamma^{(\nu}[p^{\mu)},\psi]
(\Str {\cal Y}_{j,p} \frac{1}{2} \bar{\psi'} \Gamma^{(\nu'}[p^{\mu')},\psi'])^\dagger
\;\rangle \n \\
&= \frac{1}{K^2} 
\sum_{K_1,K_2} \sum_{a,b} B_{K_1,K-K_1}^2 B_{K_2,K-K_2}^2  \n \\
&\tr {\cal Y}_1{\cal Y}_a{\cal Y}_{1'}{\cal Y}_b
(\frac{1}{P^2})^2\left(
-f P^\mu P^{\mu'} P^{\nu} P^{\nu'} \right. \n \\
&\left. +\frac{f}{8} P^2 ( P^{\mu}P^{\mu'} \delta_{\nu\nu'}+
 P^{\nu}P^{\nu'} \delta_{\mu\mu'}+
P^{\mu}P^{\nu'} \delta_{\mu'\nu}+
P^{\mu'}P^{\nu} \delta_{\mu\nu'}
) \right)
\tr {\cal Y}_b^\dagger {\cal Y}_{2'}^\dagger 
{\cal Y}_a^\dagger {\cal Y}_2^\dagger  \ ,
\end{align}
where $f(16)$ counts fermionic degrees of freedom. 
The total amplitude is obtained as
\begin{align}
A_{\rm tot}^{\mu\nu\mu'\nu'}=& \frac{1}{K^2}
\sum_{K_1,K_2} \sum_{a,b} 
B_{K_1,K-K_1}^2 B_{K_2,K-K_2}^2 \n \\
&\tr {\cal Y}_1{\cal Y}_a{\cal Y}_{1'}{\cal Y}_b
(\frac{1}{P^2})^2 
\left(
(2d-4-f) P^\mu P^{\mu'} P^{\nu} P^{\nu'} \right. \n \\
&-(1-\frac{f}{8}) P^2(P^{\mu}P^{\mu'} \delta_{\nu\nu'}
+ P^{\nu}P^{\nu'} \delta_{\mu\mu'}
+
P^{\mu}P^{\nu'} \delta_{\mu'\nu}+
P^{\mu'}P^{\nu} \delta_{\mu\nu'}
)\n \\ 
& \left.
+2 P^2  (P^{\mu}P^{\nu} \delta_{\mu'\nu'}+
 P^{\mu'}P^{\nu'} \delta_{\mu\nu}) 
+P^4 (\delta_{\mu\mu'}\delta_{\nu\nu'}+
\delta_{\mu\nu'}\delta_{\mu'\nu}) \right) 
\tr {\cal Y}_b^\dagger {\cal Y}_{2'}^\dagger 
{\cal Y}_a^\dagger {\cal Y}_2^\dagger \ , 
\end{align}
In the supersymmetric case ($f=2(d-2)$), it may be simplified further, 
\begin{align}
 A_{\rm tot}^{\mu\nu\mu'\nu'}
=&\frac{1}{K^2}
\sum_{K_1,K_2} \sum_{a,b} B_{K_1,K-K_1}^2 B_{K_2,K-K_2}^2\n \\
&\tr {\cal Y}_1{\cal Y}_a{\cal Y}_{1'}{\cal Y}_b 
\left(
\frac{2}{\tilde{d}} (\tilde{\delta}_{\mu\nu}\delta_{\mu'\nu'}
+\delta_{\mu\nu} \tilde{\delta}_{\mu'\nu'}) \right. \n \\
&  +\frac{\frac{d}{4}-\frac{3}{2}}{\tilde{d}}
(\tilde{\delta}_{\mu\mu'}\delta_{\nu\nu'}
+\tilde{\delta}_{\mu\nu'}\delta_{\mu'\nu}
+\delta_{\mu\mu'}\tilde{\delta}_{\nu\nu'}
+\delta_{\mu\nu'}\tilde{\delta}_{\mu'\nu})  \n \\
&\left. +(\delta_{\mu\mu'}\delta_{\nu\nu'}
+\delta_{\mu\nu'}\delta_{\mu'\nu})
\right) 
\tr {\cal Y}_b^\dagger {\cal Y}_{2'}^\dagger 
{\cal Y}_a^\dagger {\cal Y}_2^\dagger \ ,
\label{tot-am}
\end{align}
where we have replaced
\begin{align}
P_{\mu'}P_{\nu'}\to \frac{P^2}{\tilde{d}} \tilde{\delta}_{\mu'\nu'} \ .
\end{align}
$\tilde{d}$ denotes the dimension of the isometry group $G$.
$\tilde{\delta}_{\mu'\nu'}$ is a Kronecker delta 
in the $\tilde{d}$ dimensional subspace.

Now, let us consider the tensor structure of graviton correlators
on $ S^2 \times S^2 =SU(2)\times SU(2)/U(1)\times U(1)$. 
The dimension of $G=SU(2)\times SU(2)$ is $\tilde{d}=6$
as they can be embedded in the 6 dimensional space.
The total amplitude is obtained from (\ref{tot-am}) as
\begin{align}
& A_{\rm tot}^{\mu\nu\mu'\nu'}
=\frac{1}{K^2} \sum_{K_1} B_{K,K-K1}^4 
\left(
\frac{1}{3} (\tilde{\delta}_{\mu\nu}\delta_{\mu'\nu'}
+\delta_{\mu\nu} \tilde{\delta}_{\mu'\nu'}) \right. \n \\
&  +\frac{1}{6}
(\tilde{\delta}_{\mu\mu'}\delta_{\nu\nu'}
+\tilde{\delta}_{\mu\nu'}\delta_{\mu'\nu}
+\delta_{\mu\mu'}\tilde{\delta}_{\nu\nu'}
+\delta_{\mu\nu'}\tilde{\delta}_{\mu'\nu})
\left. +(\delta_{\mu\mu'}\delta_{\nu\nu'}
+\delta_{\mu\nu'}\delta_{\mu'\nu})
\right) \ ,
\end{align}
where we have substituted $f=16$ and $d=10$.
Tensor structure for the bosonic part is discussed in appendix (\ref{appA}).

In order to check the consistency of our calculation,
we derive the following Ward identity for the Wilson line correlators of graviton mode
\begin{align} 
&K_\mu \left(
\hspace*{1.5cm}
\begin{picture}(0,0)
\put(-40,0){\line(1,0){5}}
\put(-5,0){\line(1,0){5}}
\put(20,0){\line(1,0){5}}
\put(55,0){\line(1,0){5}}
\put(-20,0){\circle{30}}
\put(40,0){\circle{30}}
\put(-5,5){\line(1,0){30}}
\put(-5,-5){\line(1,0){30}}
\end{picture}\hspace*{8mm} 
\hspace*{15mm} 
\right) 
\n \\ \n \\
&=
K \langle\;
\Str V_{+\nu}^\dagger \Str V_{\mu'\nu'}
\;\rangle \n \\
&=
 \langle\;
(A_-)_{ij}^{K+1} (\frac{\delta}{\delta A_\nu})_{ji}
I \Str V_{\mu'\nu'} \;\rangle
-\sum_{K_1}\langle\; \tr{1\over 4}(A_-)^{K_1}
\bar{\psi}\Gamma_{-\nu}(A_-)^{K-K_1}
{\partial\over \partial \bar{\psi}} I\Str V_{\mu'\nu'} \;\rangle
\n \\
&=-\langle\;
(A_-)_{ij}^{K+1}  (\frac{\delta}{\delta A_\nu})_{ji}
\Str V_{\mu'\nu'} 
\;\rangle 
+\sum_{K_1}\langle\; \tr{1\over 4}
(A_-^{K_1}\bar{\psi}\Gamma_{-\nu}(A_-)^{K-K_1}
{\partial\over \partial \bar{\psi}}\Str V_{\mu'\nu'} \;\rangle 
\n \\
&-\delta_{-\nu } \sum_{K_1}  \langle \; \tr  
(A_-)^{K_1} \tr (A_-)^{K-K_1}
\Str V_{\mu'\nu'} 
\;\rangle \ , \label{Wardid}
\end{align} 
where 
\begin{align}
&V_{\mu\nu}=(A_+)^K \left(
[A_\rho,A_\mu][A_\rho,A_\nu]+\frac{1}{2}\bar{\psi} 
\Gamma^{(\mu} [A^{\nu)},\psi] \right) \n \\
&I={1\over 4}\Tr [A_\mu,A_\rho]^2
+\Tr\frac{1}{2}\bar{\psi} \Gamma^\mu [A_\mu,\psi] \n \\
&(A_\pm)^K=(p_\pm +a_\pm+\tilde{p}_\pm +\tilde{a}_\pm)^K
\ .
\end{align} 
These vertex operators are closely related to those we have investigated 
up to a normalization factor of $y_j^2(j!)^2/(2j)!$ since 
\begin{align}
(p_\pm+\tilde{p}_\pm)^{2j}\sim {(2j)!\over (j!)^2}
p_\pm^j\tilde{p}_\pm^j  \ .
\end{align} 
where $j>>1$ is assumed.
\footnote{The two point functions are slightly different since there are no $log(K)$ factors unlike in (\ref{logk}).}

First, let us discuss the last line in ({\ref{Wardid}}).
We focus on the leading term of the expansion of 
't Hooft coupling $\lambda$. The leading term is one loop
diagram. Therefore, the first and second trace 
can contain no creation (annihilation) operators. 
Thus, this three point function 
is calculated as 
\begin{align}
\sum_{K_1}  \langle \; \tr  
(A_-)^{K_1} \tr (A_-)^{K-K_1}
\Str V_{\mu'\nu'} 
\;\rangle 
&= 0 \ ,
\end{align}
since
\begin{align}
\tr  
 \langle \;(A_-)^K\;\rangle = 0 \quad {\rm for} \ K \neq 0 \ .
\end{align}
The one point function of Wilson line operators is 
\begin{align}
&\langle\;
-(A_-)_{ij}^{K+1}  (\frac{\delta}{\delta A_\nu})_{ji}
\Str V_{\mu'\nu'} 
\;\rangle 
+\sum_{K_1}\langle\;\tr{1\over 4}
(A_-)^{K_1}\bar{\psi}\Gamma_{-\nu}(A_-)^{K-K_1}
{\partial\over \partial \bar{\psi}}\Str V_{\mu'\nu'} \;\rangle  \ .
\end{align}
The bosonic part is calculated as
\begin{align}
& ({\cal Y}_K)^2\langle\;(A_-)^{K+1}_{ij} (\frac{\delta}{\delta A_\nu })_{ji}
\Str (A_+)^K
  [A^{\rho'},A^{\mu'}]
 [A^{\rho'},A^{\nu'}] \;\rangle\n \\
&= \frac{1}{K}\sum_{K_1} \sum_{a,b} ({\cal Y}_K)^2 \n \\
&\tr \left(
(A_{\rho'} (A_-)^{K+1} (A_+)^{K_1} f^\dagger_{\rho' \nu'} (A_+)^{K-K_1}-
A_{\rho'} (A_+)^{K_1} f^\dagger_{\rho' \nu'}  (A_+)^{K-K_1}(A_-)^{K+1}
 )\delta_{\nu\mu'} \right. \n \\
&+(A_{\rho'} (A_-)^{K+1} (A_+)^{K_1} f^\dagger_{\rho' \mu'} (A_+)^{K-K_1}-
A_{\rho'} (A_+)^{K_1} f^\dagger_{\rho' \mu'}  (A_+)^{K-K_1}(A_-)^{K+1}
 )\delta_{\nu\nu'}  \n \\
&+((A_-)^{K+1}A_{\mu'}  (A_+)^{K_1} f^\dagger_{\rho' \nu'} (A_+)^{K-K_1}-
A_{\mu'}  (A_-)^{K+1}(A_+)^{K_1} f^\dagger_{\rho' \nu'} (A_+)^{K-K_1}
 )\delta_{\rho'\nu}  \n \\ 
 &+ \left. 
((A_-)^{K+1}A_{\nu'}  (A_+)^{K_1} f^\dagger_{\rho' \mu'} (A_+)^{K-K_1}-
A_{\nu'} (A_-)^{K+1}(A_+)^{K_1} f^\dagger_{\rho' \mu'}  (A_+)^{K-K_1}
 )\delta_{\rho'\nu} \right) \ .
\end{align}
Note that if we consider the noncommutative flat space,
there are additional terms which come from the variation 
of external momenta $e^{ikA}$.
We can show that such terms do not contribute to the correlator in this regularization.

The first line of the trace part is calculated as
\begin{align}
&\frac{1}{K} \sum_{K_1} \sum_{a,b} ({\cal Y}_K)^2 \n \\
&\langle\;\;\tr
(A_{\rho'} (A_-)^{K+1} (A_+)^{K_1} f^\dagger_{\rho' \nu'} (A_+)^{K-K_1}-
A_{\rho'} (A_+)^{K_1} f^\dagger_{\rho' \nu'}  (A_+)^{K-K_1}(A_-)^{K+1}
 )\delta_{\nu\mu'}  \;\rangle \n \\
=&-\frac{1}{K^2}\sum_{K_1,K_2} \sum_{a,b}B_{K_1,K-K_1} B_{K_2,K-K_2} \n \\
& \tr {\cal Y}_1{\cal Y}_a{\cal Y}_{1'}{\cal Y}_b 
\frac{1}{P^2} 
( (d-2) P_{\nu'}K\cdot P+ P^2 K_{\nu'}) \delta_{\nu\mu'} 
\tr {\cal Y}_b^\dagger {\cal Y}_{2'}^\dagger 
{\cal Y}_a^\dagger {\cal Y}_2^\dagger
 \ .
\end{align}
In this way, the bosonic part is obtained as
\begin{align}
 & ({\cal Y}_K)^2 
\langle\;-(A_-)^{K+1}_{ij} (\frac{\delta}{\delta A_\nu })_{ji}
\Str (A_+)^K
  [A^{\rho'},A^{\mu'}]
 [A^{\rho'},A^{\nu'}] \;\rangle\n \\
\sim &\frac{1}{K^2}\sum_{K_1,K_2} \sum_{a,b}
B_{K_1,K-K_1} B_{K_2,K-K_2}  
\tr {\cal Y}_1{\cal Y}_a{\cal Y}_{1'}{\cal Y}_b  \n \\
&\frac{1}{P^2}\Big(
((d-2)K\cdot P P_{\nu'}+ P^2 K_{\nu'}) \delta_{\nu\mu'}+
((d-2)K\cdot P P_{\mu'}+ P^2 K_{\mu'}) \delta_{\nu\nu'}  \n \\
&-K\cdot P (\delta_{\mu'\nu} P_{\nu'}-P_\nu \delta_{\mu'\nu'})
+ P_{\mu'} (P_{\nu'} K_\nu-P_\nu K_{\nu'}) \n \\
& -K\cdot P (\delta_{\nu'\nu} P_{\mu'}-P_\nu \delta_{\nu'\mu'})
+ P_{\nu'} (P_{\mu'} K_\nu-P_\nu K_{\mu'}) \Big)
\tr {\cal Y}_b^\dagger {\cal Y}_{2'}^\dagger 
{\cal Y}_a^\dagger {\cal Y}_2^\dagger \ .
\end{align}
The fermionic part is calculated as
\begin{align}
&({\cal Y}_K)^2 \langle\;-(A_-)^{K+1}_{ij}  (\frac{\delta}{\delta A_\nu })_{ji}
\Str (A_+)^K
\frac{1}{2}\bar{\psi} \Gamma^{(\nu'}
[A^{\mu' )},\psi] \;\rangle\n \\
\to &\frac{1}{K^2}
\sum_{K_1,K_2} \sum_{a,b} B_{K_1,K-K_1} B_{K_2,K-K_2}
\tr {\cal Y}_1{\cal Y}_a{\cal Y}_{1'}{\cal Y}_b \n \\
&\frac{ f}{4P^2} 
(P\cdot K P_{\nu'} \delta_{\nu\mu'}+P\cdot K P_{\mu'} \delta_{\nu\nu'})
\tr {\cal Y}_b^\dagger {\cal Y}_{2'}^\dagger 
{\cal Y}_a^\dagger {\cal Y}_2^\dagger \ .
\end{align}
The contribution corresponding to fermionic equation of motion is
\begin{align}
&({\cal Y}_K)^2 \sum_{K_1}\langle\; \tr{1\over 4}
(A_-)^{K_1}\bar{\psi}\Gamma_{-\nu}(A_-)^{K-K_1}
{\partial\over \partial \bar{\psi}}\Str V_{\mu'\nu'} \;\rangle 
\n \\
\to & \frac{1}{K^2} \sum_{K_1,K_2} \sum_{a,b} B_{K_1,K-K_1} B_{K_2,K-K_2} 
\tr {\cal Y}_1{\cal Y}_a{\cal Y}_{1'}{\cal Y}_b
 \n \\
&\frac{f}{8 P^2} \left((P_\nu 
(P_{\mu'}K_{\nu'}+P_{\nu'}K_{\mu'})-
P\cdot K P_{\nu'} \delta_{\nu\mu'}-P\cdot K P_{\mu'} \delta_{\nu\nu'}\right)
\tr {\cal Y}_b^\dagger {\cal Y}_{2'}^\dagger 
{\cal Y}_a^\dagger {\cal Y}_2^\dagger \ .
\end{align}
The leading contribution of 
the one point function of Wilson line operators is given by
\begin{align}
&\frac{1}{K^2}\sum_{K_1,K_2} \sum_{a,b} B_{K_1,K-K_1} B_{K_2,K-K_2}
\tr {\cal Y}_1{\cal Y}_a{\cal Y}_{1'}{\cal Y}_b  \n \\
&\frac{1}{P^2}
\left(
(d-3-\frac{f}{4}-\frac{f}{8})(P_{\nu'} \delta_{\nu\mu'}
+P_{\mu'}\delta_{\nu\nu'})K \cdot P +2 P_\nu \delta_{\mu'\nu'}K\cdot P
\right. \n \\
&\left.
+(K_{\nu'} \delta_{\nu\mu'}+K_{\mu'}\delta_{\nu\nu'})P^2
+2 K_\nu P_{\mu'} P_{\nu'}
+(\frac{f}{8}-1)P_\nu 
(P_{\mu'}K_{\nu'}+P_{\nu'}K_{\mu'})
\right)
\tr {\cal Y}_b^\dagger {\cal Y}_{2'}^\dagger 
{\cal Y}_a^\dagger {\cal Y}_2^\dagger \ .
\label{1pt}
\end{align}
By multiplying $K_\mu$ to $A_{\rm tot}^{\mu\nu\mu'\nu'}$, we obtain
\begin{align}
K_\mu A_{\rm tot}^{\mu\nu\mu'\nu'}&
=\frac{1}{K^2} \sum_{K_1,K_2} \sum_{a,b}
B_{K_1,K-K_1}
 B_{K_2,K-K_2}
\tr {\cal Y}_1{\cal Y}_a{\cal Y}_{1'}{\cal Y}_b \n \\
&{1\over P^2} \Big(
(2d-4-f) \frac{K \cdot P P^{\mu'} P^{\nu} P^{\nu'}}{P^2}
+ 2P_{\nu}\delta_{\mu'\nu'}K\cdot P+2K_{\nu}P_{\mu'}P_{\nu'}
 \n \\
&+(\frac{f}{8}-1)
((P_{\nu'} \delta_{\nu\mu'}
+P_{\mu'}\delta_{\nu\nu'})K \cdot P
+P_\nu 
(P_{\mu'}K_{\nu'}+P_{\nu'}K_{\mu'}))  \n \\
& +P^2(K_{\nu'} \delta_{\nu\mu'} +K_{\mu'}\delta_{\nu\nu'})
\Big) 
\tr {\cal Y}_b^\dagger {\cal Y}_{2'}^\dagger 
{\cal Y}_a^\dagger {\cal Y}_2^\dagger \ .
\label{ward2}
\end{align}
When $f=2d-4$, (\ref{ward2}) and (\ref{1pt}) agree with each other.

\section{Universality of the result\label{s3}}

\subsection{Universal amplitude
\label{s31}}

As we have seen in the previous section, the 
Wilson line correlator is given by the separating function 
$B_{j_1,j-j_1}$. In this section,
we will show that this result is universal since it only assumes 
the completeness condition of the generators of $SU(N)$.

The correlators contain the following amplitude
\begin{align}
\hspace*{1.5cm}
\begin{picture}(0,0)
\put(-40,0){\line(1,0){5}}
\put(-5,0){\line(1,0){5}}
\put(20,0){\line(1,0){5}}
\put(55,0){\line(1,0){5}}
\put(-20,0){\circle{30}}
\put(40,0){\circle{30}}
\put(-5,5){\line(1,0){30}}
\put(-5,-5){\line(1,0){30}}
\put(-17,0){${\cal Y}_1'$}
\put(5,10){${\cal Y}_a$}
\put(-47,4){${\cal Y}_1$}
\put(5,-15){${\cal Y}_b$}
\end{picture}\hspace*{8mm} 
\hspace*{15mm} =
\tr {\cal Y}_1{\cal Y}_a  {\cal Y}_{1'} {\cal Y}_b
\tr {\cal Y}_b^\dagger {\cal Y}_{2'}^\dagger 
{\cal Y}_a ^\dagger {\cal Y}_2^\dagger
 \ . 
\end{align} \vspace*{0.1cm}

We recall the completeness condition:
\begin{align} \label{comp}
\sum_a &({\cal Y}_a)_{ij} ({\cal Y}_a^\dagger)_{kl}=\delta_{il} \delta_{jk} \ .
\end{align}
By using this relation, we obtain
\begin{align}
&\sum_{ab} \tr  {\cal Y}_1{\cal Y}_a
 {\cal Y}_{1'}{\cal Y}_b
\tr {\cal Y}_b^\dagger {\cal Y}_{2'}^\dagger
{\cal Y}_a^\dagger  {\cal Y}_2^\dagger \n \\
=& \tr {\cal Y}_1 {\cal Y}_2^\dagger
 \tr {\cal Y}_{1'} {\cal Y}_{2'}^\dagger \ .
\end{align}
While ${\cal Y}$ depends on a particular $G/H$ we pick,
the following relation is universal
\begin{align}
\tr {\cal Y}_1 {\cal Y}_2^\dagger
=\delta_{j_1-j_2}+{\cal O}(1/N) \ ,
\end{align}
where $j_1$ is a momentum carried by ${\cal Y}_1$.
$\tr {\cal Y}_{1'} {\cal Y}_{2'}^\dagger$ provides the same $\delta$ 
due to the momentum conservation law.
The universality of the amplitude
reflects on the universality with respect to 
the topology of the D-brane worldvolume, which 
is closely related to the cut off independence of the analysis.

\begin{figure}[hbtp]
\epsfysize=4cm
\begin{center}
\epsfbox{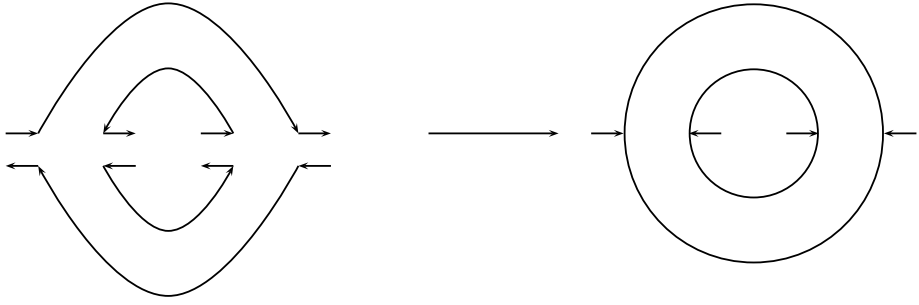}
\end{center}
\caption{'t Hooft's double line notation. 
We sum over the internal momenta which constitute a complete set 
of states.}
\label{thooft}
\end{figure}

Finally, we provide a pictorial representation of
our evaluation of the universal amplitude in figure 
\ref{thooft}. 
Our result is naturally understood by using the 
't Hooft's double line notation. 

\subsection{Example : $CP^2$ \label{s32}}

In contrast to the preceding sub-section, the separating function $B$ depends on 
a choice of $G/H$.
In this sub-section, we will show that the momentum ($k$) dependence of 
WLC on 
$CP^2=SU(3)/U(2)$ is also as $1/k^2$.
We will calculate $B$ in the semiclassical 
approximation.
We define the raising and lowering operators as
\begin{align}
p_{\pm}=\frac{1}{\sqrt{2}}(p_4\pm ip_5) \ , \quad
\tilde{p}_{\pm}=\frac{1}{\sqrt{2}}(p_6\pm ip_7) \ ,
\end{align}
 
The normalization condition of spherical harmonics is
\begin{align}
\tr {\cal Y}_j^\dagger {\cal Y}_j=1 \ ,
\end{align}
where 
\begin{align}
{\cal Y}_j=y_j (p_+)^j \ .
\end{align}
In the semiclassical approximation, 
\begin{align}
p_{+}=r \frac{\xi_1}{1+\bar{\xi} \xi} \ , \quad
p_- =r\frac{\xi_2}{1+\bar{\xi}{\xi}} \ ,
\end{align}
we may estimate
\begin{align}
\tr {\cal Y}_j^\dagger {\cal Y}_j&=
r^{2 j+2} \int \frac{2 d^4 \xi}{\pi^2 (1+\xi\bar{\xi})^3}
\frac{(\bar{\xi}\xi)^{j}}{(1+\bar{\xi}\xi)^{2j}}
y_j^2 \n \\
&=r^{2j+2}\frac{2(j!)^2}{(2j+2)!}y_j^2 \ .
\end{align}
Thus, we obtain 
\begin{align}
\tilde{B}_{j_1,j-j_1}^2&=\frac{y^2_j}{y^2_{j-j_1}y^2_{j_1}} \n \\
&\sim \frac{\sqrt{\pi}}{2} (\frac{j}{(j-j_1)j_1})^{3\over 2}N \ .
\end{align}
The Wilson
line correlators (\ref{strsum}) are calculated as
\begin{align}
&\langle\; \Str {\cal Y}_{j} f_1 f_2
\Str f_2^\dagger f_1^\dagger {\cal Y}_{j}^\dagger
\;\rangle \n \\
&=  \frac{1}{j^2} \sum_{j_1=0}^j  \sum_{j_2=0}^j 
 \langle\; \tr {\cal Y}_{j_1} f_1 {\cal Y}_{j-j_1} f_2
\tr f_2^\dagger {\cal Y}_{j_2}^\dagger f_1^\dagger 
{\cal Y}_{j-j_2}^\dagger
\;\rangle B_{j_1j-j_1} B_{j_2j-j_2}
\n \\
&\sim \frac{N}{j^2} \sqrt{\pi} \zeta(\frac{3}{2}) \ .
\end{align}
We have obtained the $1/({\rm momentum})^2$ behavior without a $\log$ factor.
The correlators are also invariant under the rotation of 8
dimensional space in which $CP^2$ sits.

\subsection{Universality with respect to the dimensionality
\label{s33}}

We have shown in this section that the 
correlator is given by the separating function $B$.
This result holds for any $G/H$ , irrespective of its dimension. 
Therefore, we consider higher dimensional NC gauge theory here.
NC gauge theory on $S^2 \times S^2 \times S^2$ is 
considered in \cite{fuzS2S2S2}.
The WLC is obtained as 
\begin{align}
&\langle\; \Str (p_{a+})^j (p_{b+})^j (p_{c+})^j f_1 f_2
\Str f_2^\dagger f_1^\dagger (p_{a-})^j (p_{b-})^j (p_{c-})^j
\;\rangle  \n \\
&=  \frac{n^2 }{j^2} \sum_{j_1=0}^j 
({y_j\over y_{j_1}y_{j-j_1}})^6 \n \\
& =  \frac{n^2 }{j^2} \sum_{j_1=0}^j 
B_{j_1,j-j_1}^6  \n \\
&\sim  \frac{n^2 }{j^2} \sum_{j_1=1}^j  
(\frac{l^2 \pi j}{j_1(j-j_1)})^{\frac{3}{2}} \n \\
&=\frac{N n \pi^{3/2}}{4 j^2} \zeta (\frac{3}{2})\ .
\end{align}
Thus, the graviton is localized on 6 dimensional subspace: 
$S^2 \times S^2 \times S^2$.
We may naturally interpret that graviton is localized on D5-brane.

When we consider  $(S^2 \times)^x$ type spacetime, 
correlators are calculated as
\begin{align}
\frac{n^2 }{ j^2} \sum B^{2x}
\sim N\frac{n}{j^2} \ .
\end{align}
except $S^2 (x=1)$.
Thus, the correlators exhibit the inverse squared momentum law
on any $G/H$ whose dimension is larger than $2$.

\section{Conclusions and Discussions \label{s4}}

In this paper, we have investigated the two point correlation functions of
graviton vertex operators in 4 dimensional NC gauge theory with maximal SUSY 
on compact homogeneous spacetime $G/H$.
The infrared contributions ($k^4log(k)$) to the correlators are identical to
those in conformal field theory  just like
the correlators of the energy-momentum tensor.
However the ultra-violet contributions are very different even in the small
external momentum case. 
This is due to the UV/IR mixing effects caused by the NC phases in the correlators.
In the case of the symmetric ordered graviton operators,
we find that the two point correlators behave as $1/k^2$.
This fact indicates the existence of massless gravitons in NC gauge theory.
It has been clear that there is a bulk gravity in 4d NC gauge theory 
with maximal SUSY since the one loop effective action involving
the quadratic Wilson lines is consistent with 10 dimensional supergravity.
In order to obtain realistic quantum gravity, we need to obtain 4 dimensional gravity.
Such a possibility may be realized in various ways if a graviton is bound to the brane
or through induced gravity on the brane.
We hope our findings will make a first concrete step to identify such a
mechanism in 4d NC gauge theory.

We still need to investigate various issues to establish such a mechanism.
One issue is to understand the correlators of $n$ point functions.
Another issue is to understand the correlators of more generic Wilson lines.
If we consider the vertex operators which contain more commutators
of $[A_{\mu},A_{\nu}]$, analogous calculations show that 
the two point functions are more singular in the infra-red limit 
than $1/k^2$.  
It might imply that the relevant modes are (gravitationally) confined and 
develop a mass gap in that channel.
On the other hand, the correlators of the Wilson lines 
which contain fewer $[A_{\mu},A_{\nu}]$
do not exhibit singularity in the infra-red limit.
The third issue is that the two point correlators are not transverse
due to the one point functions as we have seen in the Ward identity. 
They seem to correspond to graviton propagators in a certain gauge.

Our investigation is also restricted to the leading order of the 't Hooft coupling
in NC gauge theory
which is valid in the weak coupling regime. We need to understand higher
order quantum corrections also. Since the behavior of the correlators
is governed by the power counting, it is likely that higher order 
corrections do not
modify our results. It is also desirable to have a consistent supergravity
description in the strong coupling limit.

If the graviton vertex operators are coupled to conserved energy-momentum
tensor, we can reproduce the Newton's law between them by taking
the expectation values of the graviton vertex operators.
It might be a good strategy to pursue this idea further since
such a structure is consistent with the one loop effective action
of NC gauge theory.

\acknowledgments

This work is supported in part by the Grant-in-Aid for Scientific Research
from the Ministry of Education, Science and Culture of Japan.
The work of S. N. is supported in part by Research Fellowships of 
the Japan Society for the Promotion of Science for Young Scientists. 

\appendix

\section{Bosonic part of the tensor structure of graviton correlators on 
$S^2 \times  S^2$ \label{appA}}

In this appendix, 
we investigate the bosonic part of the tensor structure of graviton
correlators on $S^2 \times S^2$. We obtain the anisotropic tensor
structure. For the supersymmetric correlators, we obtain the isotropic
tensor structure in section \ref{s23}. By considering the isometry of
the space, we can replace
\begin{align}
P_\mu P_\nu P_\mu' P_\nu' &\to
\frac{P_A^4}{15}(\delta_A^{\mu\nu}\delta_A^{\mu'\nu'}+
\delta_A^{\mu\mu'}\delta_A^{\nu\nu'}+
\delta_A^{\mu\nu'}\delta_A^{\mu'\nu}) \n \\
&+\frac{P_A^2P_B^2}{9}(\delta_A^{\mu\nu}\delta_B^{\mu'\nu'}+
\delta_A^{\mu\mu'}\delta_B^{\nu\nu'}+
\delta_A^{\mu\nu'}\delta_B^{\mu'\nu})\n \\
&+\frac{P_A^2P_B^2}{9}(\delta_B^{\mu\nu}\delta_A^{\mu'\nu'}+
\delta_B^{\mu\mu'}\delta_A^{\nu\nu'}+
\delta_B^{\mu\nu'}\delta_A^{\mu'\nu})\n \\
&+\frac{P_B^4}{15}(\delta_B^{\mu\nu}\delta_B^{\mu'\nu'}+
\delta_B^{\mu\mu'}\delta_B^{\nu\nu'}+
\delta_B^{\mu\nu'}\delta_B^{\mu'\nu}) \ , \label{repl}
\end{align}
where $\delta_A$ and $\delta_B$ are Kronecker delta
effective to the 3 dimensions, 
\begin{align}
&\delta_A^{\mu\nu}=
\left(\begin{array}{cccccc}
1&&&&&\\
&1&&&&\\
&&1&&&\\
&&&0&&\\
&&&&0&\\
&&&&&0
\end{array}\right),
\delta_B^{\mu\nu}=
\left(\begin{array}{cccccc}
0&&&&&\\
&0&&&&\\
&&0&&&\\
&&&1&&\\
&&&&1&\\
&&&&&1
\end{array}\right) \n \\
&P_A^2=P_4^2+P_5^2+P_6^2, \quad
P_B^2=P_7^2+P_8^2+P_9^2 \ .
\end{align}
By using (\ref{repl}), the bosonic part of the correlator 
(\ref{tens1}) is replaced as
\begin{align}
&(\frac{16}{15}P_A^4+\frac{2}{3}P^4)(\delta_A^{\mu\nu}\delta_A^{\mu'\nu'}+
\delta_A^{\mu\mu'}\delta_A^{\nu\nu'}+
\delta_A^{\mu\nu'}\delta_A^{\mu'\nu}) \n \\
+&(\frac{16}{9}P_A^2P_B^2+\frac{2}{3}P^4)
(\delta_A^{\mu\nu}\delta_B^{\mu'\nu'}+
\delta_A^{\mu\mu'}\delta_B^{\nu\nu'}+
\delta_A^{\mu\nu'}\delta_B^{\mu'\nu}) \n \\
+&(\frac{16}{9}P_A^2P_B^2+\frac{2}{3}P)
(\delta_A^{\mu\nu}\delta_B^{\mu'\nu'}+
\delta_A^{\mu\mu'}\delta_B^{\nu\nu'}+
\delta_A^{\mu\nu'}\delta_B^{\mu'\nu}) \n \\
+&(\frac{16}{15}P_B^4+\frac{2}{3}P^4)(
\delta_B^{\mu\nu}\delta_B^{\mu'\nu'}+
\delta_B^{\mu\mu'}\delta_B^{\nu\nu'}+
\delta_B^{\mu\nu'}\delta_B^{\mu'\nu}) \ . \label{isotro}
\end{align}
We need to estimate the $P_A^4$ and $P_A^2P_B^2$.
We calculate them under the 
semiclassical approximation.
Angular momenta are represented by the adjoint representation 
on $S^2$, then, the integral of $P_A^4$ is semiclassically written as
\begin{align} \label{semicl}
\int  dX_1 d\tilde{X}_1 \frac{(X_1-X_2)^4}{(X_1-X_2)^2
+(\tilde{X_1}-\tilde{X_2}^2)}
\end{align}
where
\begin{align}
P_A =X_1-X_2 \ .
\end{align}
$X_2$ and $\tilde{X}_2$ are fixed at some point on $S^2$.
(\ref{semicl}) is calculated as
\begin{align}
&\int d\Omega d\tilde{\Omega}  \frac{(X_1-X_2)^4}{(X_1-X_2)^2
+(\tilde{X_1}-\tilde{X_2}^2)} \n \\
=&\int_0^\pi d \cos \theta d \cos \tilde{\theta}
\frac{(2-2 \cos^2 \theta)^2}{(4-2 \cos^2 \theta 
-2 \cos^2 \tilde{\theta})^2} \n \\
=&\int_{-1}^1 d X d \tilde{X}\frac{(1-X^2)^2}{(2-X^2-\tilde{X}^2)^2}
\ ,
\label{pa4}
\end{align}
where we transform the valuables as
\begin{align}
X=\cos \theta, \quad
\tilde{X}=\cos \tilde{\theta} \ .
\end{align}
The integral of $P_A^2P_B^2$ is also estimated as
\begin{align} \label{papb}
\int_{-1}^1 d X d \tilde{X}
\frac{(1-X^2)(2-\tilde{X}^2)}{(2-X^2-\tilde{X}^2)^2} \ .
\end{align}
By carrying out the integration of $\tilde{X}$ in (\ref{papb}),
 we obtain
\begin{align}
4 \int_0^1 dX (-1+X^2) \left(-\frac{-1+X^2}{2 (-2+X^2)(-1+X^2)}
+\frac{(-3+X^2)\tan^{-1}\frac{1}{\sqrt{-2+X^2}}}{2 (-2+X^2)^{3/2}}
\right) \ .
\end{align}
The first term is calculated as
\begin{align} \label{firstt}
-2+\sqrt{2}\log(1+\sqrt{2}) \ .
\end{align}
The second term is calculated as
\begin{align} \label{second}
-4 \int_1^{\sqrt{2}} d x \frac{(x^2+1)(x^2-1)}{2 x^2 \sqrt{2-x^2}}
\tanh^{-1}\frac{1}{x} \ ,
\end{align}
where we transform the valuables as
\begin{align}
X^2-2=-x^2 \ .
\end{align}
Formally, $\tanh (1/x) $ is expanded as
\begin{align}
\tanh^{-1}\frac{1}{x}=
\sum_{k=1}^\infty \frac{1}{2k-1} (\frac{1}{x})^{2k-1} \ .
\end{align}
By using this expression, we carry out the integral in (\ref{second}) as
\begin{align} 
\sum_{k=1}^\infty \frac{-4}{2k-1} \left(
-\frac{2^{-5/2-k}}{k(k-2)} (\frac{(k-2)\sqrt{\pi} \Gamma (1-k)}{\Gamma
 (1/2-k)}-\frac{4k\sqrt{\pi} \Gamma (3-k)}{\Gamma (5/2-k)})
\right.
 \n \\ \left.
+\frac{-(k-2) _2 F_1 ((1/2,1),(1-k),-1)+k _2 F_1((1/2,1),(3-k),-1)
}{4k(k-2)}
\right) \ , \label{secondt}
\end{align}
where $_2 F_1(a;b;z)$ is a generalized hypergeometric function. 
We numerically obtain (\ref{papb}) 
as
\begin{align}
\int_{0}^1 d X d \tilde{X}
\frac{(1-X^2)(2-\tilde{X}^2)}{(2-X^2-\tilde{X}^2)^2}
&\sim-0.188+0.396 \n \\
&=0.208 \ . \label{num1}
\end{align}
We also evaluate (\ref{pa4}) as
\begin{align}
\int_{0}^1 d X d \tilde{X}\frac{(1-X^2)^2}{(2-X^2-\tilde{X}^2)^2}
\sim 0.292 \ . \label{num2}
\end{align}
After all, the Bosonic part of the tensor structure of the graviton 
on $S^2 \times S^2$ (\ref{isotro}) 
is evaluated among the estimations (\ref{num1}) and (\ref{num2}) as
\begin{align}
&0.98(\delta_A^{\mu\nu}\delta_A^{\mu'\nu'}+
\delta_A^{\mu\mu'}\delta_A^{\nu\nu'}+
\delta_A^{\mu\nu'}\delta_A^{\mu'\nu}+
\delta_B^{\mu\nu}\delta_B^{\mu'\nu'}+
\delta_B^{\mu\mu'}\delta_B^{\nu\nu'}+
\delta_B^{\mu\nu'}\delta_B^{\mu'\nu}) \n \\
+&1.04
(\delta_A^{\mu\nu}\delta_B^{\mu'\nu'}+
\delta_A^{\mu\mu'}\delta_B^{\nu\nu'}+
\delta_A^{\mu\nu'}\delta_B^{\mu'\nu}+
\delta_A^{\mu\nu}\delta_B^{\mu'\nu'}+
\delta_A^{\mu\mu'}\delta_B^{\nu\nu'}+
\delta_A^{\mu\nu'}\delta_B^{\mu'\nu})  \ .
\end{align}


\begin{thebibliography}{99}

\bibitem{CDS} A. Connes, M. Douglas and A. Schwarz,
{\em Noncommutative Geometry and Matrix Theory:
Compactification on Tori},
\jhep{02}{1998}{003}, \hepth{9711162}.
\bibitem{AIIKKT}
H. Aoki, N. Ishibashi, S. Iso, H. Kawai, Y. Kitazawa
and T. Tada,
{\em Non-commutative Yang-Mills in IIB Matrix Model},
\npb{565}{2000}{176},
\hepth{9908141}.
\bibitem{Li}
M. Li, {\em Strings from IIB Matrices},
\npb{499}{1997}{149}, \hepth{961222}.
\bibitem{IKKT}N. Ishibashi, H. Kawai, Y. Kitazawa and A. Tsuchiya,
{\em A Large-N Reduced Model as Superstring},
\npb{498}{1997}{467}, \hepth{9612115}.
\bibitem{BFSS}
T. Banks, W. Fischler, S.H. Shenker and L. Susskind,
{\em M-theory As A Matrix Model: A Conjecture},
\prd{55}{1997}{5112}, \hepth{9610043}.
\bibitem{MRS}
S. Minwalla, M.V. Raamsdonk, N. Seiberg,
{\em Noncommutative Perturbative Dynamics},
\jhep{02}{2000}{020},
\hepth{9912072}.
\bibitem{Susskind}
L. Susskind,
{\em The Anthropic Landscape of String Theory},
\hepth{0302219}.
\bibitem{Maldacena}
J. Maldacena, {\em The Large $N$ Limit of Superconformal Field theories
and Supergravity}, \atmp{2}{1998}{231},
\hepth{9711200}.
\bibitem{RS}
L. Randall and R. Sundrum,
{\em An Alternative to Compactification},
\prl{83}{1999}{4690}, \hepth{9906064}.
\bibitem{KTT2}
Y. Kitazawa, Y. Takayama and D. Tomino,
{\em Wilson Line Correlators in N=4 Non-commutative Gauge Theory on $S^2
	\times S^2$}, \npb{715}{2005}{665},
\hepth{0412312}.
\bibitem{fuzS2}
T. Imai, Y. Kitazawa, Y. Takayama and D. Tomino,
{\em Quantum Corrections on Fuzzy Sphere},
\npb{665}{2003}{520},
\hepth{0303120}.
\bibitem{fuzS2S2}
T. Imai, Y. Kitazawa, Y. Takayama and D. Tomino,
{\em Effective Actions of Matrix Models on
Homogeneous Spaces},
\npb{679}{2004}{143},
\hepth{0307007}.
\bibitem{KTT1}
Y. Kitazawa, Y. Takayama and D. Tomino,
{\em Correlators of Matrix Models on
Homogeneous Spaces},
\npb{700}{2004}{183},
\hepth{0403242}.
\bibitem{fuzS2S2S2}
H. Kaneko, Y. Kitazawa and D. Tomino,
{\em Stability of Fuzzy $S^2 \times S^2 \times S^2$ in IIB Type Matrix Models},
\npb{725}{2005}{93},
\hepth{0506033}.
\bibitem{fuzCP2}
H. Kaneko, Y. Kitazawa and D. Tomino,
{\em Fuzzy Spacetime with SU(3) Isometry in IIB Matrix Model},
\hepth{0510263}.
\bibitem{Myers}
R.C. Myers,{\em Dielectric-Branes},
\jhep{12}{1999}{022}, \hepth{9910053}.
\bibitem{Mathom}
Y. Kitazawa,
{\em Matrix Models in Homogeneous Spaces},
\npb{642}{2002}{210},
\hepth{0207115}.
\bibitem{Edm}
A. R. Edmonds,
{\em Angular Momentum in Quantum Mechanics},
Princeton Univ. Press (1957).
\bibitem{IIKK}
N. Ishibashi, S. Iso, H. Kawai and Y. Kitazawa, {\em Wilson Loops in
Non-commutative Yang-Mills}, \npb{573}{2000}{573},
\hepth{9910004}.
\bibitem{Gross}
D. J. Gross, A. Hashimoto and N. Itzhaki,
{\em Observables of Non-Commutative Gauge Theories},
\atmp{4}{2000}{893}, \hepth{0008075}.
\bibitem{DhKhe}
A. Dhar and Y. Kitazawa,
{\em High-Energy Behavior of Wilson Lines},
\jhep{02}{2001}{004}, 
\hepth{0012170}.
\bibitem{vertex}
Y. Kitazawa,
{\em Vertex Operators in IIB Matrix Model},
\jhep{04}{2002}{004},
\hepth{0201218}.
\bibitem{ITU}
S. Iso, H. Terachi and H. Umetsu,
{\em Wilson Loops and Vertex Operators in Matrix Model},
\prd{70}{2004}{125005},
\hepth{0410182}.
\bibitem{VMK}
D.A. Varshalovich, A.N. Moskalev and V.K. Khersonskii,
{\em Quantum theory of angular momentum : irreducible tensors, 
spherical harmonics, vector coupling coefficients, 3nj symbols},
World Scientific, 1988.

\end{thebibliography}
\end{document}